\documentclass[
    ,final            
    ,numberedheadings 
   ,cmfonts
  ]
  {aipproc}

\layoutstyle{6x9}

\usepackage{amsmath}

\newcommand{\vect}[1]{\vec{#1}}

\newcommand{\eqs}[1]{\begin{eqnarray}#1\end{eqnarray} }
\newcommand{\mvec}[1]{|\vec{#1}|}
\newcommand{\matrice}[1]{\begin{matrix} #1\end{matrix} }
\newcommand{\ks}[1]{#1 \!\!\! \slash } 

\newcommand{\nn}{\nonumber}

  {\begin{displaymath}%
      \everymath={\displaystyle}%
      \begin{array}{l}}%
  {\end{array}\end{displaymath}}

\begin{document}

\title{Pion Structure Function and Violation of the Momentum Sum Rule}

\author{J.P.~Lansberg\footnote{email: JPH.Lansberg@ulg.ac.be}, F.~Bissey, J.R.~Cudell, J.~Cugnon, M.~Jaminon and P.~Stassart
}{
  address={Universit\'e de  Li\`ege, D\'epartement de  Physique B5, Sart Tilman,B-4000 LIEGE 1, Belgium}
}

\begin{abstract}

We present a method to evaluate the pion structure functions from
a box diagram calculation. Pion and
constituent quark fields are coupled through the simplest pseudoscalar
coupling. The $\gamma^{\star} \pi \rightarrow q \bar q$ 
cross-section is evaluated and related to the structure functions. 
We then show that the introduction of non-perturbative
effects, related to the pion size and preserving gauge invariance, provides us 
with a straighforward relation with the quark distribution. It is predicted that
higher-twist terms become negligible for $Q^2$ larger than about 2~GeV$^2$ and
that quarks in the pion have a momentum fraction smaller than in the proton. 
We enlarge the discussion concerning this violation of the momentum sum rule, 
emphasizing that the sum rule is recovered in the chiral limit 
and also when the finite size condition is not imposed.

\end{abstract}

\maketitle


\section{Introduction}

The usual way to describe quarks inside hadrons relies on 
the structure functions.
Quantum Chromodynamics (QCD), being the theory of the interactions between
quarks and gluons, is supposed to predict the structure 
functions, but in fact, these quantities result from 
non-perturbative effects and hence are not easily modeled. 
QCD is only able to predict their evolution as the virtuality 
of the probing photon ($Q^2$) changes.
It is our purpose here to build a simple model for these quantities.

Phenomenological quark models, which possess some non-perturbative
aspects and which are rather successful in reproducing low-energy
properties of hadrons, are expected to  help us 
understand the connection between deep-inelastic 
scattering (DIS) data and non-pertubative
inputs. For very simple systems such as pions and other low-mass
mesons, there exist some effective models ({\it e.g.} the 
Nambu-Jona-Lasinio (NJL) model) that
incorporate, in some simplified way, special QCD features 
such as spontaneous chiral symmetry breaking and anomalies.

However,  investigations of the structure functions along these lines
have given rather different
results~\cite{SH93,DA95,Davidson:2001cc}. This situation originates from the fact that the
NJL model
needs to be regularized and that different regularizations yield different
results. 

We present here an alternative way to evaluate of the pion structure function 
based on Ref.~\cite{Bissey:2002yr}, where
the $q \overline{q} \pi$ vertex is represented by the simplest
pseudoscalar coupling, {\it i.e.} $i g \gamma_5$ and the pion size 
is mimicked by the introduction of a gauge preserving cut-off.

\section{Cross-Section Calculation for $\gamma^\star \pi_0 \to q \bar q$ by Integration over $t$}
\label{sec:gpiqq}

In this section, we calculate the  cross-section for $\gamma^\star \pi_0 \to q \bar q$ 
which can be related to the form factors $W_1$ and $W_2$ as we shall see later. Another 
way to achieve this goal consists in calculating the imaginary part of the forward amplitude for  
$\gamma^\star \pi_0 \to \gamma^\star \pi_0$. 

\begin{figure}

\begin{minipage}[t]{7.5cm}
\hfill
{\includegraphics[width=3.0cm]{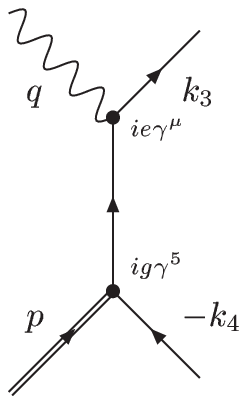}}
\end{minipage}
\begin{minipage}[t]{4.0cm}
\
\end{minipage}
\begin{minipage}[t]{7.5cm}
{\includegraphics[width=3.0cm]{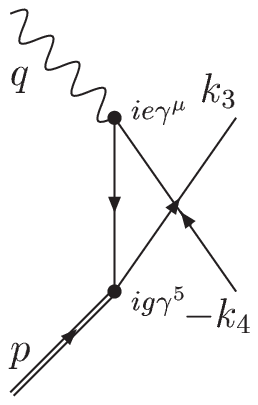}}
\end{minipage}
 \caption{Feynman diagrams for the process $\gamma^\star \pi_0 \to q \bar q$.}\label{fig:gpiqq}  
\end{figure}

At the leading order in the loop expansion, we have the two diagrams shown in Fig.~\ref{fig:gpiqq} 
which, when squared, give two types of contributions. The first one (corresponding to box-diagrams 
for the $\gamma^\star \pi_0 \to \gamma^\star \pi_0$ amplitude) enables us to calculate 
the structure functions straightforwardly. The second involves crossed quark lines and is related 
to processes where the two photons (in the $\gamma^\star \pi_0 \to \gamma^\star \pi_0$ scattering) 
are connected to different quark lines. This is incompatible with a probabilistic 
interpretation in terms of structure functions. As we will conclude 
from the calculations, contributions of this kind are fortunately 
higher-twists, {\it i.e.} they scale as $\frac{1}{Q^2}$, when we introduce the cut-off. 

Their introduction is nevertheless mandatory, as this ensures gauge invariance 
(see~Ref.~\cite{Bissey:2002yr}). This is required for a proper evaluation of the 
cross-section for the  $\gamma^\star \pi_0 \to q \bar q$ process, which is a
physical process, observable and thus gauge invariant. 

\subsection{Kinematics of the Process}\label{subsec:gpiqq_kin}

For the sake of convenience, we work with cartesian coordinates in the 
CM (center of momentum) frame. The four-momenta of the four particles are given by:
\eqs{\label{eq:four_momenta}
q= (E(=q_0), \vec q),
p=(E_\pi,-\vec q),   
k_3=(k_{3,0},\vec k_3),
k_4=(k_{4,0},-\vec k_3). 
}

Let us now introduce the Mandelstam variables:
\eqs{
&&s=(p+q)^2=(k_3+k_4)^2= E_{CM}^2=(E+E_\pi)^2, \nn\\
&&t=(k_3-q)^2=(k_4-p)^2,u=(k_3-p)^2=(k_4-q)^2 .
}
We then define the square momenta:
\eqs{
&&q^2=-Q^2, p^2=m^2_\pi, k_3^2=m^2, k_4^2=m^2  }

The variables of Eq.~(\ref{eq:four_momenta}) can be written  in terms of these quantities
and of Mandelstam variables. One has:
\eqs{ 
k_{3,0}=k_{4,0}=\frac{\sqrt{s}}{2}&,& \mvec{k_3}=\mvec{k_4}=\sqrt{\frac{s}{4}-m^2}, }

\eqs{
E&=&\frac{s-m_\pi^2-Q^2}{2\sqrt{s}},}
and
\eqs{
\mvec{p}^2=\mvec{q}^2=E^2+Q^2=\frac{(s-m_\pi^2-Q^2)^2+4sQ^2}{4s}. 
}

\begin{figure}

\includegraphics[width=5cm]{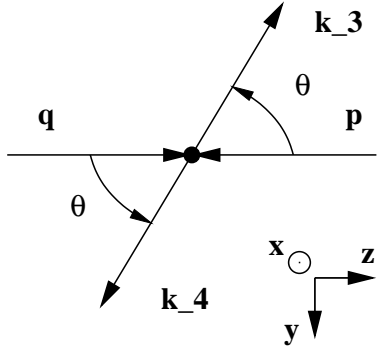}
  \caption{Kinematics and definition of the angle $\theta$ in the CM frame.}
\label{fig:cinematique_GammaPion}
\end{figure}

The angle $\theta$ between the vectors $\vect k_3$ and $\vect q$ 
(see Fig.~\ref{fig:cinematique_GammaPion}) is related to $t$ by 
\eqs{
t
&=&m^2+m_\pi^2 -s+\sqrt{s} \frac{s-m_\pi^2-Q^2}{2\sqrt{s}} +2\sqrt{\frac{s}{4}-m^2}
\frac{\sqrt{(s-m_\pi^2-Q^2)^2+4sQ^2}}{2\sqrt{s}}\cos\theta\nn
}
Eventually we obtain: 
\eqs{\label{eq:cos_theta}\cos\theta=\frac{t-m^2-\frac{m_\pi^2}{2} +\frac{s}{2}+
\frac{Q^2}{2}}{2\sqrt{\frac{s}{4}-m^2}\frac{\sqrt{(s-m_\pi^2-Q^2)^2+4sQ^2}}{2\sqrt{s}}}=
\frac{t-m^2-\frac{m_\pi^2}{2} +\frac{s}{2}+\frac{Q^2}{2}}{2\sqrt{\frac{s}{4}-m^2}\sqrt{E^2+Q^2}}.}

The physical values of $t$ are lying between boundaries, corresponding to $\cos\theta=\pm1$:
\begin{equation}
\label{eq:t_min,max}t_{\tiny \matrice{\rm max \\\rm  min}}= m^2+\frac{m_\pi^2}{2} 
-\frac{s}{2}-\frac{Q^2}{2} \pm 2\sqrt{\frac{s}{4}-m^2}\frac{\sqrt{(s-m_\pi^2-Q^2)^2+4sQ^2}}{2\sqrt{s}}
\end{equation}

Finally, assuming the incident direction along the $z$ axis and the scattering in the $y-z$ plane, we get:
\eqs{\label{eq:momenta_gpiqq}
q&=& (\frac{s-m_\pi^2-Q^2}{2\sqrt{s}},0,0,\sqrt{\frac{(s-m_\pi^2-Q^2)^2+4sQ^2}{4s}})=(E,0,0,\sqrt{E^2+Q^2}),  \nn\\
p&=&(\sqrt{m_\pi^2+\frac{(s-m_\pi^2-Q^2)^2+4sQ^2}{4s}},0,0,-\sqrt{\frac{(s-m_\pi^2-Q^2)^2+4sQ^2}{4s}})\nn\\
k_3&=&(\frac{\sqrt{s}}{2},0,-\sqrt{\frac{s}{4}-m^2}\sin(\theta),\sqrt{\frac{s}{4}-m^2}\cos\theta), \nn\\
k_4&=&(\frac{\sqrt{s}}{2},0,\sqrt{\frac{s}{4}-m^2}\sin(\theta),-\sqrt{\frac{s}{4}-m^2}\cos\theta). 
}
These quantities will be explicitly used in the determination of scalar products involving the polarisation
vectors.

\subsection{Amplitude Calculation}\label{sec:gpiqq_calc_ampl}

Applying Feynman rules for the processes shown in Fig.~\ref{fig:gpiqq}, we obtain 
the analytical expression of the square amplitude in a straightforward way. 
For a given polarisation of the virtual photon, $i=L,T_1,T_2$, and with well-known 
conventions\footnote{The isospin/charge factor $(e_u^2+e_d^2)$ corresponds to the following 
choice of the isospin matrix: $\pi^-:\left(\matrice{ 0 & 0 \\ \sqrt{2} &0}\right)$ -- $\pi^0:\left(\matrice{ 1 & 0\\ 0 &-1}\right)$
-- $\pi^+:\left(\matrice{ 0 &   \sqrt{2}\\ 0 &0}\right)$
-- $\gamma:\left(\matrice{ e_u&  0 \\ 0&e_d}\right)$  } we have for the squared of the first diagram 
shown in Fig.~\ref{fig:gpiqq},
\eqs{\label{eq:ampl_gpiqq}
\left|{\cal M}_i\right|^2= 
3 g^2 (e_u^2+e_d^2) 
{\rm Tr}&\hspace*{-0.6cm}& \left(  (-\ks k_4+m)\gamma^5\frac{\ks k_3-\ks q+m}{(k_3-q)^2-m^2}\gamma^{\mu'} \right.\nn\\ 
&\hspace*{-0.6cm}&\left. (\ks k_3+m) \gamma^\mu \frac{\ks k_3-\ks q+m}{(k_3-q)^2-m^2}\gamma^5 
 \right)\varepsilon_{i\mu} \varepsilon^\star_{i\mu'} .
}
The expression for the square of the second diagram as well as for the interference terms have similar 
expressions.

We first proceed to the calculation of the expressions of the scalar products of the 
polarisation vectors with $q, p, k_3$ and $k_4$ for the three possible polarisation states.


We shall use the following definition of the longitudinal polarisation vector, 
\eqs{
\varepsilon_{L,\mu}=\frac{1}{\sqrt{Q^2}}(q_z,0,0,E).
}

One has
\eqs{
\varepsilon_L\cdot q &=& 0, \nn\\
\varepsilon_L\cdot p &=& \frac{1}{\sqrt{Q^2}} \sqrt{s} \sqrt{E^2+Q^2} \ 
, \nn\\
\varepsilon_L\cdot k_3 &=& \frac{1}{\sqrt{Q^2}} \left(\frac{\sqrt{s}\sqrt{E^2+Q^2}}{2}-\frac{E(t-m^2-\frac{m_\pi^2}{2} +\frac{s}{2}+\frac{Q^2}{2})}{2\sqrt{E^2+Q^2}}\right).
}


For the linear transverse polarisation vectors, we use the following definition:
\eqs{
\varepsilon_{T_1,\mu}&=&(0,1,0,0), \ \varepsilon_{T_2,\mu}=(0,0,1,0). 
}

This and Eq.~(\ref{eq:momenta_gpiqq}) lead to:
\eqs{
\varepsilon_{T_i}\cdot q &=&0 \ = \ \varepsilon_{T_i}\cdot p   , \ \ i=1,2,\nn\\
\varepsilon_{T_1}\cdot k_3 &=& 0, \nn\\
\varepsilon_{T_2}\cdot k_3 &=& \sqrt{\frac{s}{4}-m^2}
\sqrt{1-\left(\frac{t-m^2-\frac{m_\pi^2}{2} +\frac{s}{2}+\frac{Q^2}{2}}{2\sqrt{\frac{s}{4}-m^2}
\sqrt{E^2+Q^2}}\right)^2}.
}

\subsection{Polarised Cross-Sections}

The total cross-section is given by the following relation~\cite{barger}, 
with $i=L,T_1,T_2$:
\eqs{\frac{{\rm d}\sigma_i}{{\rm d}t}=\frac{1}{16\pi\lambda}\left|{\cal M}_i\right|^2}
with\footnote{The function $\lambda(x,y,z)$ is defined as  
$\lambda(x,y,z)\equiv x^2+y^2+z^2-2xy-2xz-2yz$} $\lambda=\lambda(s,-Q^2,m^2_\pi)$. 
The latter directly links the amplitude --which is obtained, in our case, with REDUCE-- 
 and the cross-section differential upon $t$. Integrating
over $t$ from $t_{\rm min}$ to $t_{\rm max}$ 
(see Eq.~(\ref{eq:t_min,max})) gives the total cross-sections for 
given polarisations:
\eqs{\label{eq:sigma_i}
\sigma_{i}= \frac{1}{16\pi\lambda} 3 g^2 (e_u^2+e_d^2) \int \!{\rm d}t \ 
T^{\mu\mu'} \varepsilon_{i,\mu} \varepsilon^\star_{i\mu'} 
,}
with $T^{\mu\mu'}\equiv{\rm Tr} \left(
(-\ks k_4+m)\gamma^5\frac{\ks k_3-\ks q+m}{(k_3-q)^2-m^2}\gamma^{\mu'}(\ks k_3+m)
\gamma^\mu \frac{\ks k_3-\ks q+m}{(k_3-q)^2-m^2}\gamma^5 
\right)$.

It is also of interest to consider 
\eqs{\label{eq:sigma_contr}
\sigma_{contr.}=\frac{1}{16\pi\lambda}3g^2 (e_u^2+e_d^2) \int \!{\rm d}t \ 
T^{\mu\mu'} 
(g_{\mu\mu'}+\frac{q_\mu q_{\mu'}}{Q^2})
}
since it does not involve any polarisation vectors and its calculation is somewhat easier. However,
it should not confused with $\sigma_L+\sigma_{T_1}+\sigma_{T_2}$ (see later), which could be measured
in some limiting experimental conditions.

\subsection{The Cut-Off}\label{sec:perf_integration}

\subsubsection{Cut-off and Higher-twists}

We introduce a cut-off in our calculations: 
instead of integrating over $t$ from $t_{\rm min}$ to $t_{\rm max}$, we integrate 
from\footnote{Recall that $t<0$.}
$Max\{t_{\rm min},\Lambda_t\}$ to $t_{\rm max}$. As shown
in Ref.~\cite{Bissey:2002yr}, this is equivalent to setting a cut-off for the square
of the relative momentum of the quarks inside the pion, which can be considered as 
resulting from the finite size of the pion.

\begin{figure}[b]
\includegraphics[width=10cm]{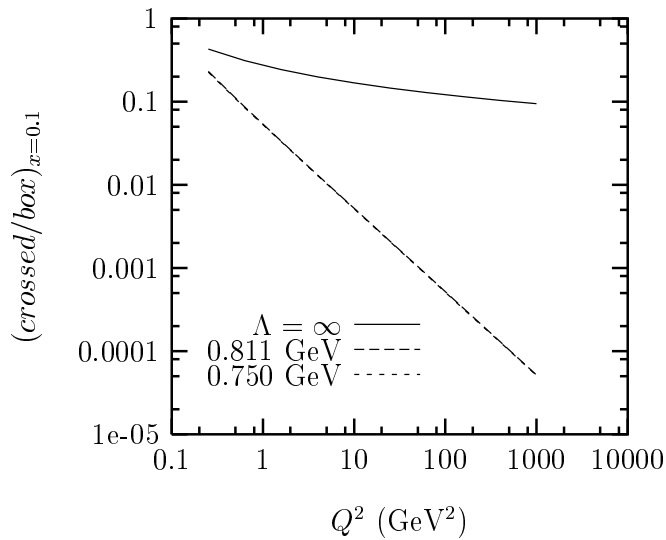}
  \caption{Ratio of the direct contributions to the crossed contributions to the cross-section for
$\gamma^\star \pi^0\to q \bar q$ at $x=0.1$ and for three values of the cut-off $\Lambda$.}
\label{fig:higher_twist_proof}
\end{figure}

Irrespectively of the factors arising in the extraction of the structure functions (see 
section~(\ref{sec:sfnp_sig&w1})), we can simply convince ourselves that in the presence of a cut-off 
the contribution of the crossed diagrams (the first photon hits a quark and the second an 
antiquark) are higher-twists, by plotting the ratio of these two contributions as a function 
of $Q^2$. The suppression is 
clear on Fig.~\ref{fig:higher_twist_proof} for whatever (finite) value of the cut-off; the curves
are even indistinguishable in this case.

As we already said, in order to get a physical interpretation of our results 
--in terms of structure functions--, 
we need the crossed diagram contributions to be suppressed. The importance
of a cut-off, which is not a regulator in the studied process, is thus manifest.

\subsubsection{Cut-off and Gauge Invariance}

In most cases, the presence of a cut-off implies the breakdown of gauge invariance. This is the main
reason why one commonly uses other regularization methods such as Pauli-Villars or dimensional 
regularization. In the context of an effective theory, a simple momentum 
cut-off has the critical advantage that the regularization
parameter (which is kept to a finite value) can be interpreted as a physical one. In the case of 
dimensional regularization, the signification of fractional dimensional space is not at all clear.
As for Pauli-Villars regularization, the introduction of wrong-sign fermionic contribution is not
intuitive.

Nevertheless, in our model, as the intermediate states in 
$\gamma^\star \pi^0 \to \gamma^\star \pi^0$ are on-shell, the cut-off on the 
internal loop momentum\footnote{which corresponds, up to some constants, to a cut-off on the 
relative mometum between the two quarks at the $\pi^0q \bar q$ vertex.} acts on a physical 
observable, that is $t$ of the process 
$\gamma^\star \pi^0 \to q \bar q$.
The conclusion is thus staightforward: this cut-off does not break gauge invariance. 

This procedure would have been fully justified if we had only one pion vertex, and thus one 
condition to implement. But we have two vertices with two different
expressions for the relative momenta. To keep gauge invariance, the cut-off on $t$ must be on the overall
cross-section as the physical observable is the full cross-section (the square of the sum of the 
two diagrams) at a given $t$.

To keep the constraint coming from the pion size, we should have imposed two conditions on the 
cross-section to be integrated over $t$. However, this would have given zero. As a makeshift we 
imposed one condition or the other.

Even if the physical interpretation is not so obvious, we thus kept gauge invariance, information
from the pion size, suppressed the crossed diagrams and hence obtained an interpretation in terms
of structure functions.

\section{$\sigma_{\gamma^\star \pi_0 \to q \bar q}$, $W_1$, $W_2$, $F_1$ and $F_2$}
\label{sec:sfnp_sig&w1}

At this stage we have calculated the polarised cross-sections and we should be able to link them 
to the structure functions for neutral pion. Before doing this, we shall first 
recall some mandatory formulae derived from the analysis of deep inelastic proton
scattering. 

We introduce the usual form of the hadronic tensor
\eqs{\label{eq:wmunu1}W_{\mu\nu}(p,q)= (-g_{\mu\nu}-\frac{q_\mu q_\nu}{Q^2}) W_1(Q^2,\nu)+
[p_\mu+(\frac{p\cdot q}{Q^2})q_\mu]
[p_\nu+(\frac{p\cdot q}{Q^2})q_\nu]\frac{W_2(Q^2,\nu)}{M^2},}
with $M$, the hadron mass, and 
$W_1(Q^2,\nu)$ and $W_2(Q^2,\nu)$
the structure functions.

In the same spirit as for the definition of the cross-section of
$\gamma^\star p \to X$ \cite{aitchison},
we define the polarised cross-sections for  $\gamma^\star \pi_0 \to X$ by
\footnote{This definition is independent of the hadron spin because, on the one hand, we have to 
sum the contributions coming from the different polarisations and on the other hand, we have 
a spin factor arising in the relation between $W_{\mu\nu}$ and the hadronic current.}:

\eqs{\sigma_i(\gamma^\star \pi_0 \to X)=\frac{4\pi^2\alpha}{K}\varepsilon^{\star\mu }_i 
\varepsilon^{\nu}_i W_{\mu\nu}, 
} 
with
the flux factor $K=\frac{1}{2}\frac{\sqrt{\lambda(s,-Q^2,m^2_\pi)}}{m_\pi}$.

Defining the transverse cross-section through
\eqs{
\sigma_T\equiv\frac{1}{2}(\sigma_{T_1}+\sigma_{T_2}),
}
we can easily show that:
\begin{eqnarray} 
\sigma_T&=&\frac{4\pi^2\alpha}{K}W_1, \label{eq:w_1}\nn\\
\sigma_L&=&\frac{4\pi^2\alpha}{K}\left[(1+\frac{\nu}{2xm_\pi^2})W_2-W_1\right]. \label{eq:w_1&2}
\end{eqnarray}

These last two formulae establish the link between the functions $W_1$ and $W_2$ and 
previously calculated polarised cross-sections for 
$\gamma^\star \pi_0 \to q \bar q$~(Eq.~(\ref{eq:sigma_i}))\footnote{
Besides we would like to stress that one has to be cautious if one wants to consider 
non-polarised processes. Usually, 
a simple contraction of the amplitude for a given helicity with the expression, 
\eqs{-g^{\mu\nu}+\frac{q^\mu q^\nu}{M^2}} seems to be the most convenient way to proceed, 
instead of considering separately the different polarisations. This is in most cases licit 
because
\eqs{\label{eq:sum_pol_massive}\sum_{i=L,T_1,T_2} \varepsilon^{\star\mu }_i \varepsilon^{\nu}_i
=-g^{\mu\nu}+\frac{q^\mu q^\nu}{M^2},}
for massive particles, and thus the --strict-- sum over polarisations gives the {\it correct} tensor.

For spacelike virtual particles, nevertheless, we have the following normalisations
\eqs{\label{eq:norm_pol_trans}
\varepsilon_T\cdot\varepsilon_T&=&-1, \nn\\
\varepsilon_L\cdot\varepsilon_L&=&+1.}

Therefore, the  following relation --which indeed reduces to Eq.~(\ref{eq:sum_pol_massive}) for 
massive particles for which Eq.~(\ref{eq:norm_pol_trans}) does not hold--,
\eqs{ \sum_i \frac{\varepsilon^{\star\mu }_i \varepsilon^{\nu}_i}
{\varepsilon_i\cdot\varepsilon_i}=g^{\mu\nu}+\frac{q^\mu q^\nu}{Q^2}}
leads, in this case, to 
\eqs{
\sigma_{contr.}=\frac{4\pi^2\alpha}{K}\left(g^{\mu\nu}+\frac{q^\mu q^\nu}{Q^2}\right)
W_{\mu\nu}&=&\sigma_L-2\sigma_T}
but not to $\sigma_L+2\sigma_T$ !}.

With all these tools, we can now extract the functions $W_1$ and $W_2$. 
Inverting for $W_1$ and $W_2$, we get
\eqs{W_1&=&\frac{\sqrt{\lambda}}{8\pi^2m_\pi\alpha} \sigma_T
}
and
\eqs{W_2=\frac{\sqrt{\lambda}}{8\pi^2m_\pi\alpha}\frac{2xm_\pi^2}{\nu+2xm_\pi^2}(\sigma_L+\sigma_T).}

We then get  $F_1$ and  $F_2$ through the usual scaling relations:
\eqs{\label{eq:scaling} &&F_1(x)=MW_1(Q^2,\nu)\ ,  F_2(x)=\frac{\nu}{M}W_2(Q^2,\nu)\ , 
x= \frac{Q^2}{2 p\cdot q}=\frac{Q^2}{2\nu},
} still with $M$ for the hadron mass.

\section{Sum Rules, Plots and Results}

\subsection{Number of Particles Sum Rule}

The structure functions can now be related to the (valence) quark distributions:
\begin{equation}
F_1=\frac{4}{18}(u_v(x)+\bar{u}_v(x)) + \frac{1}{18}(d_v(x)+\bar{d}_v(x)),
\label{F1}
\end{equation}
\begin{equation}
F_2=2x F_1.
\label{F2}
\end{equation}
We stress that the last relation, known as the Callan-Gross relation, comes out
of our calculation. This does indicate that our approximations have been done
consistently\footnote{For charged pions, the additional diagrams implying a
direct coupling of  the virtual photon to the pion are suppressed by a factor
$1/s$ and the leading-twist results are the same, except for the charge
coefficients entering Eq.~(\ref{F1}) .}.
\par
\begin{figure}
{\includegraphics[width=10cm]{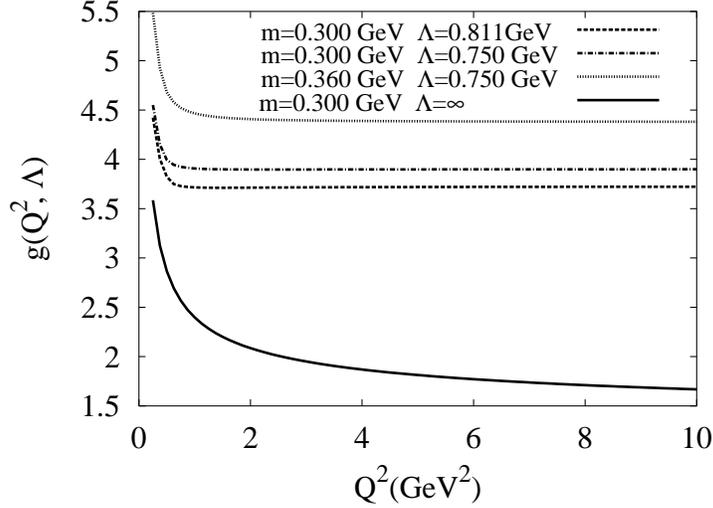}}
\caption{Values of the coupling constant $g(Q^2,\Lambda)$ which fulfill  the sum
rule (Eq.~(\ref{normF1})), for the values of the parameters indicated at the
top.}
\label{fig:f2}
\end{figure}

So far, we have only described the model. In order to make predictions, we need
to fix its parameters, namely \( \Lambda  \), \( m \) and \( g \). The
latter can be thought of as the normalisation of the quark wave function, and
we determine it by imposing that there are only two constituent quarks in the
pion. In our model, the valence quark distributions are equal
\begin{equation}
u_v(x)=\bar{u}_v(x)=d_v(x)=\bar{d}_v(x)\equiv v(x).
\label{v}
\end{equation}
The condition \( \int _{0}^{1}v(x)dx=1/2 \) then yields
\begin{equation}
\label{normF1}
\int _{0}^{1}F_{1}(x)dx=\frac{5}{18}.
\end{equation}
 As \( F_{1} \) is a function, not only of \( m \) and \( \Lambda  \),
but also of \( Q^{2} \), this gives us a coupling constant that evolves with
\( Q^{2} \). The resulting values of \( g \) are shown in Fig.~\ref{fig:f2} 
(right): for a finite
cut-off \( \Lambda  \), the cross-section at fixed \( g \) would grow with
energy, until the pion reaches its maximum allowed size, in which case the
cross
section would remain constant. If we impose relation Eq.~(\ref{normF1}),
this means that \( g(Q^{2}) \) will first decrease until the cut-off makes
it reach a plateau for \( Q^{2}\gg \Lambda ^{2} \) (in practice, the
plateau is reached around \( Q^{2} \approx 2\Lambda ^{2} \)). The plateau
value depends on \( m/\Lambda  \) (and \( m_{\pi }/\Lambda  \)).

\subsection{Momentum Sum Rule}

To constrain further our parameters, we can try to use the momentum sum rule
\begin{equation}
\label{mf}
2\left<x\right>=4\int _{0}^{1}xv(x)dx=\frac{18}{5}\int _{0}^{1}F_{2}(x)dx.
\end{equation}
In the parton model, this integral should be equal to 1 as \( Q^{2}\rightarrow
\infty  \),
as we do not have gluons in the model.

\begin{figure}[h]
{\includegraphics[width=12cm]{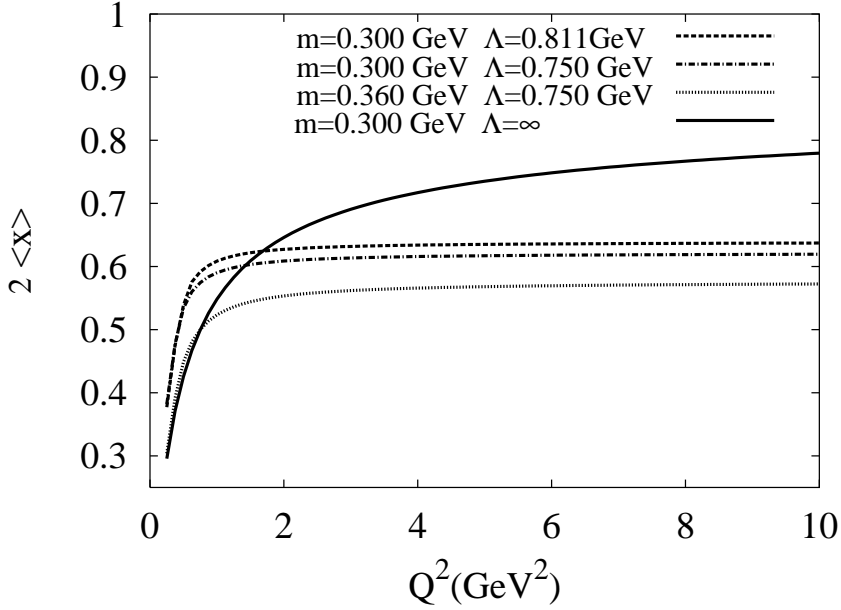}}
\caption{Momentum fraction of the quarks inside the neutral pion
(Eq.~\ref{mf}), as a function of  $Q^2$, for the values of the parameters
indicated at the top of the figure.}
\label{fig:f4}
\end{figure}

Analyzing Fig.~\ref{fig:f4}, it is clear that -- at least for a finite value
of $\Lambda$-- the momentum sum rule is violated. For instance the momentum
carried by the quarks at large  $Q^2$ is $0.6$ for $m=300$ MeV and $\Lambda=750$ MeV.

Nevertheless, there are two limits of
our model that fulfill the condition \( 2\left<x\right>=1 \) automatically and for 
which the quarks carry the entire momentum of the pion. First of all, if
we do not impose that the pion has a finite size, then asymptotically we get
\( 2\left<x\right>= \)\( 1 \). In the limit \( m_{\pi }=0 \), we obtain
the simple expression
\begin{equation}
\label{mfch}
2\left<x\right>(\Lambda =\infty , m_{\pi }=0)=\frac{4\ln (2\nu /m^{2})-3}{4\ln (2\nu
/m^{2})-1}.
\end{equation}
This means that in that regime, for sufficiently high \( Q^{2} \), the quarks
behave as free particles, and the usual derivation based on the OPE holds
\cite{IT}. 

The second case where this holds confirms this interpretation: if we impose 
\break \(
m\leq m_{\pi }/2 \),
the expressions we have given develop an infrared divergence, which corresponds
to the case where both quarks emerging from the pion are on-shell and free.
This divergence can be re-absorbed into the normalisation Eq.~(\ref{normF1})
of \( g \), and the sum rule \( 2\left<x\right>=1 \) is again automatically satisfied at
large $Q^2$. 

However, in the physical pion case, it makes more sense to consider that one of the
quarks remains
off-shell: the introduction of a cut-off changes the sum rule value,
as the fields can never be considered as free. 
Hence, because of the Goldstone
nature
of the pion, one expects that the momentum sum rule will take a smaller value
than in the case of other hadrons. This may explain why fits that assume
the same momentum fraction for valence quarks in protons and pions \cite{GRS}
do not seem to leave any room for sea quarks \cite{Zeus}.

\begin{figure}[h]
{\includegraphics[width=12cm]{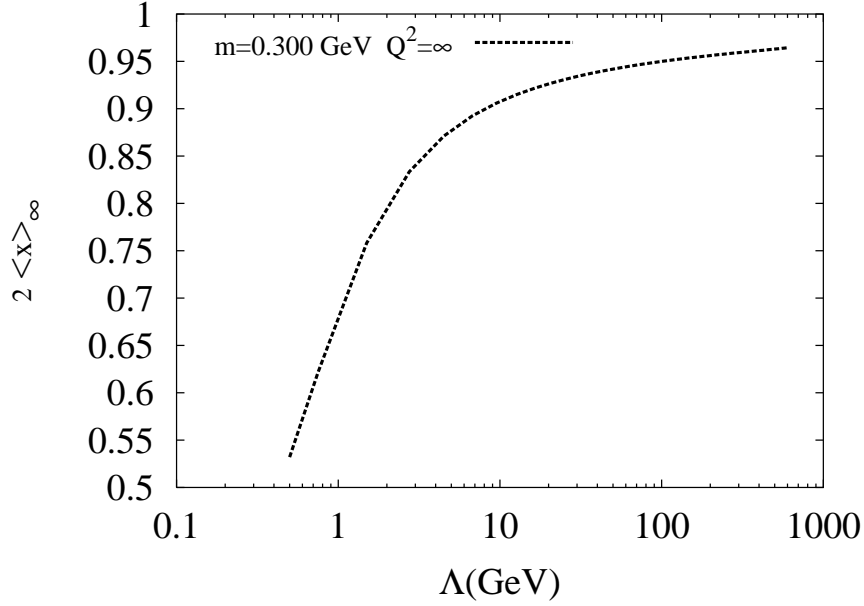}}
\caption{Asymptotic value (for large $Q^2$) 
of the momentum fraction of the
quarks inside the neutral pion (Eq.~\ref{mf}), as a function of the cut-off
$\Lambda$.}
\label{fig:f5}
\end{figure}

We come back to the results for \( 2\left<x\right> \) shown in Fig.~\ref{fig:f4}. 
Again, the curves show a plateau at sufficiently large
\( Q^{2} \),
with a value depending on \( m/\Lambda  \) and \( m_{\pi }/\Lambda  \).
It is easy to show that this value is always smaller than 1 as it can be guessed
from Fig.~\ref{fig:f5}.

Following the  discussion in~\cite{Bissey:2002yr}, we choose the following 
conservative values for the remaining parameters of our model: \( m=300 \)~MeV
or \( 360 \)~MeV and \( \Lambda \simeq 800 \)~MeV.

\subsection{Plots of $F_2$}
\begin{figure}[h]
\begin{minipage}[t]{7.5cm}
{\includegraphics[width=7.5cm]{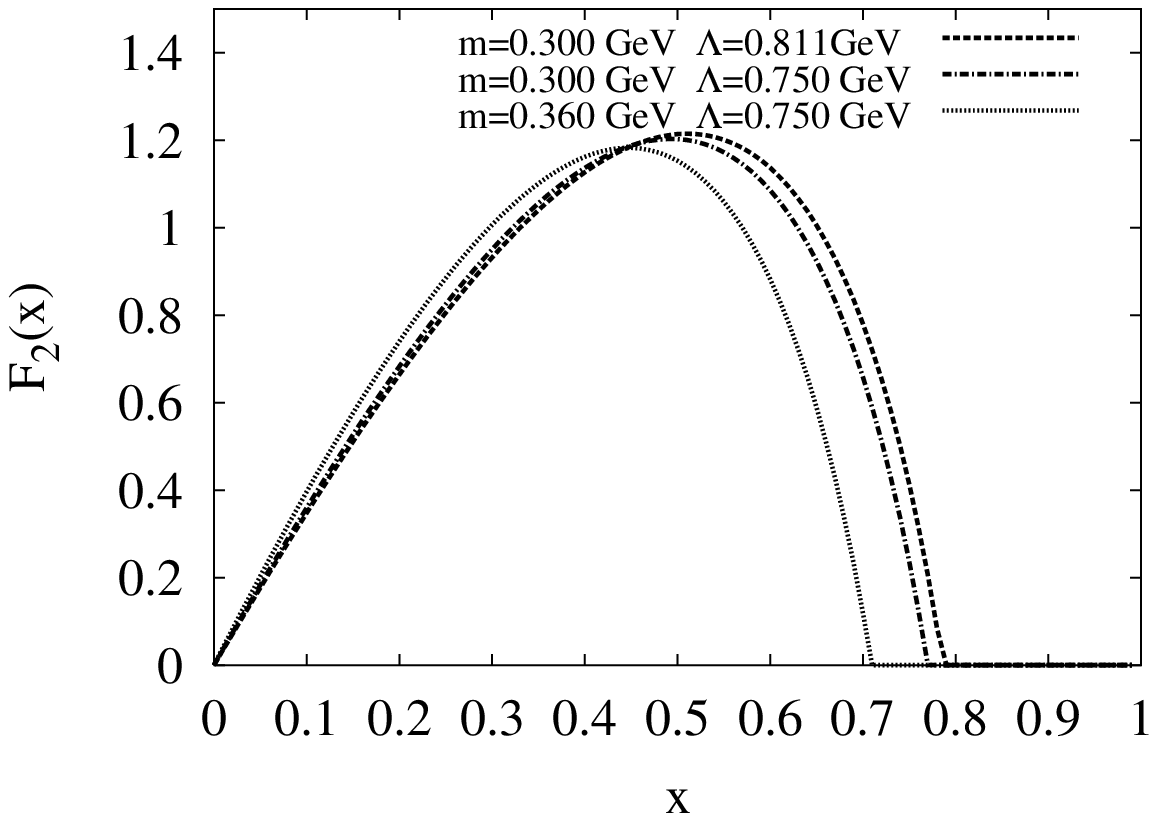}}

\end{minipage}
\hfill
\begin{minipage}[t]{7.5cm}
{\includegraphics[width=7.5cm]{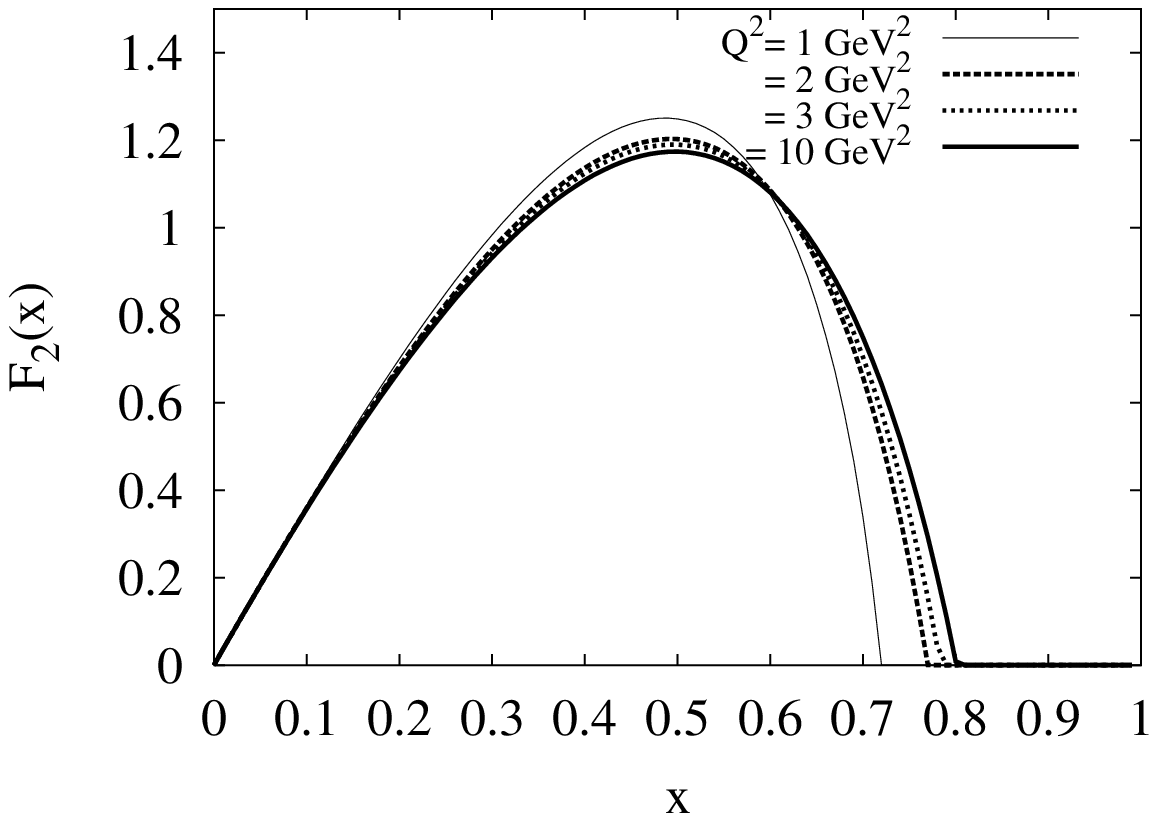}}

\end{minipage}
\caption{Structure function $F_2$ for the neutral pion. On the left,
$Q^2$=2~GeV$^2$ and the values of the parameters $m$ and $\Lambda$ are as
indicated. On the right, $m$=0.3~GeV and  $\Lambda$=0.75~GeV and the values
of $Q^2$ are as indicated.}
\label{f6}

\end{figure}

Let us finally examine the properties of the distribution $v(x)$,
or equivalently, of the function $F_2$. Some of our results are summarized in
Fig.~\ref{f6}, for  \break $Q^2$ =  2~GeV$^2$. The most striking feature is the
vanishing of this function for $x$ larger than some value $x_{max}$. This is
again due to the fact  that, because the  quarks are not free in this model,
the actual value of their mass does matter. It can  clearly be seen that this
effect  originates from kinematical cuts (see Ref.~\cite{Bissey:2002yr}).

As a result,
putting the cut quarks on-shell will be easier for small  than for large $x$.
Therefore, the $x$-distribution ($F_1$) is expected to be enhanced on the low
$x$ side, leading to a momentum fraction smaller than unity.  The vanishing at
large $x$ is not obtained in similar works, in particular in the one of 
Ref.~\cite{SH93}. In this reference, the Bjorken limit is taken first and the
kinematical constraint 
is not applied. This procedure seems
questionable for evaluating  cross-sections at finite $Q^2$. 

\section{Conclusion}
We have discussed the simplest model allowing to relate  
$\gamma^\star \pi \to q \bar q$ to the pion quark distributions. The model
consists in calculating the simplest diagrams for a $\gamma^5$ coupling
between pion and constituent quark fields.

We show that the crossed diagrams, which
are to be included in order to guarantee gauge invariance, preclude the 
existence of the relation between the the cross-section and the quark 
distribution. The introduction of a cut-off on $t$ allows such a relation
without breaking gauge invariance. The cut-off on $t$ being equivalent to a
cut-off on the square of the relative momentum of the quarks inside pion, 
this effect can be traced back to the finite size of the pion. For reasonable
value of the cut-off, the crossed diagrams become negligible as soon as $Q^2$ 
is larger than 2 GeV$^2$ allowing quark distributions to be meaningful even 
for relatively small $Q^2$. 

One of the most important outcomes of our calculation is the deviation from the 
momentum sum rule. However, we recover this sum rule ($2\left<x\right>=1$) in
two limiting cases, which correspond to free quarks. The first one is the limit
$\Lambda \to \infty$. The second one is for $m < \frac{m_\pi}{2}$.
Then, an infrared divergence appears, which can be reabsorbed into the normalisation
of $g$ and the momentum sum rule is again recovered for large $Q^2$. Hence, 
the violation of the
sum rule can tentatively be attributed to the Goldstone boson nature of the pion.


\begin{theacknowledgments}
This work has been performed in the frame of ESOP Collaboration and 
has benefited from the financial support of the EU (contract N$^\circ$ HPRN-CT-2000-00130).
\end{theacknowledgments}

\end{document}